\documentclass[sn-aps,Numbered]{sn-jnl}

\usepackage{graphicx}%
\usepackage{multirow}%
\usepackage{amsmath,amssymb,amsfonts}%
\usepackage{amsthm}%
\usepackage{mathrsfs}%
\usepackage[title]{appendix}%
\usepackage{xcolor}%
\usepackage{textcomp}%
\usepackage{manyfoot}%
\usepackage{booktabs}%
\usepackage{algorithm}%
\usepackage{algorithmicx}%
\usepackage{algpseudocode}%
\usepackage{listings}%

\begin{document}

\title[Radiative Corrections in Super Rosenbluth Experiments]{Radiative Corrections in Super Rosenbluth Experiments}

\author[1]{\fnm{Quinn} \sur{Stefan}}

\author*[1]{\fnm{Axel} \sur{Schmidt}}\email{axelschmidt@gwu.edu}

\affil[1]{\orgdiv{Department of Physics}, \orgname{The George Washington University}, \orgaddress{\street{725 21st St.\ NW}, \city{Washington}, \state{DC} \postcode{20052}, \country{USA}}}

\abstract{
Super Rosenbluth experiments, elastic electron-proton scattering experiments that eschew traditional electron detection and opt instead for the detection of the recoiling proton, have several experimental advantages. One claimed advantage is that radiative corrections are more favorable, i.e., smaller and with less kinematic dependence. In this paper, we explore this claim by conducting Monte Carlo simulations of both Super Rosenbluth and traditional Rosenbluth experiments with different models of radiative effects. When using a model that employs the peaking approximation, we indeed confirm the reduced kinematic dependence of the radiative corrections. However, we find that more sophisticated models that avoid the peaking approximation are unable to produce numerically stable results, due to a large enhancement to the cross section for bremsstrahlung radiation from the proton when the momentum transfer, $Q^2$, approaches zero. Since this enhancement is not modeled in the peaking approximation, a more robust approach to radiative corrections in Super Rosenbluth experiments is needed.
}


\maketitle

\section{Introduction}
\label{sec:intro}

The electromagnetic form factors of the proton, $G_E$ and $G_M$, which encode the distributions of charge and magnetism, respectively, have been mapped from decades worth of elastic electron-proton scattering measurements (see, e.g., \cite{Hofstadter:1960zz,Janssens:1965kd,Bartel:1973rf,Walker:1993vj,Andivahis:1994rq,E94110:2004lsx,Qattan:2004ht,A1:2010nsl,Christy:2021snt}). 
The form factors are functions of a single variable, $Q^2$, the negative squared four-momentum transfer; $G_E^2$ and $G_M^2$ can be independently determined for a given value of $Q^2$ through the traditional technique of Rosenbluth separation. 
By making measurements of the unpolarized elastic scattering cross section with several combinations of beam energy and scattering angle, but with $Q^2$ fixed, the contributions from the electric and magnetic form factors can be disentangled. 
This technique faces several experimental challenges; it relies on multiple measurements of absolute cross sections made at different kinematics with starkly different counting rates. 
The analysis of the data invariably leads to strong anti-correlation in the inferred $G_E^2$ and $G_M^2$. 

These challenges, as well as the emergence of a discrepancy in the early 2000s between determinations of $G_E/G_M$ from polarization transfer measurements~\cite{BatesFPP:1997rpw,JeffersonLabHallA:1999epl,Gayou:2001qt,A1:2001xxy,JeffersonLabHallA:2001qqe,Punjabi:2005wq,MacLachlan:2006vw,Puckett:2010ac} and determinations from earlier Rosenbluth separations,
were part of the impetus for new experiments based on the so-called ``Super Rosenbluth'' technique. In a typical Rosenbluth separation, a spectrometer is used to detect the scattered electron. By contrast, in a Super Rosenbluth experiment, only the recoiling proton is detected. One major advantage of the Super-Rosenbluth technique is that for fixed $Q^2$, the out-going proton always has the same momentum, regardless of the initial beam energy; hence there is no need to modify the magnetic field setting of the spectrometer between measurements, reducing systematic uncertainty. 
Two Super Rosenbluth measurements were conducted at Jefferson Lab~\cite{E01-001,E05-017}, and results from one were published~\cite{Qattan:2004ht}, confirming the results of prior, traditional Rosenbluth measurements. 

Ref.~\cite{Qattan:2004ht} claims another advantage of Super Rosenbluth measurements is a reduced sensitivity to radiative effects. Specifically, Ref.~\cite{Qattan:2004ht} writes that ``radiative corrections (mainly electron bremsstrahlung) $\ldots$ have smaller $\varepsilon$-dependence when the proton is detected," where $\varepsilon$ is the virtual photon polarization parameter. This is advantageous because in Rosenbluth separation, the two form factors are disentangled based on the $\varepsilon$-dependence of the cross section. Smaller $\varepsilon$-dependence of extraneous effects, such as radiative corrections, means a reduced risk of bias in the results. Ref.~\cite{Qattan:2004ht} goes on to write that for the experiment in question, ``Radiative corrections to the cross section are 20\%, with a 5\%--10\% $\varepsilon$-dependence, smaller than in previous Rosenbluth separations where the electron was detected.''

The possible advantage of proton detection in terms of reducing sensitivity to radiative effects comes from the fact that bremsstrahlung radiation is primarily emitted from the electron (either incoming or outgoing) owing to the electron's much smaller mass. By ignoring the electron and detecting the proton, one need not model the precise shape of the so-called radiative tail of the electron as it loses energy to radiation. This argument seems sound until considering that calculating the radiative correction to a proton detection experiment requires one to integrate over all possible out-going electron momenta. The electron is a light particle, prone to radiation, and one must consider this radiation not only when the emitted energy is small, but also when it is very large. 

In this paper, we revisit the claims about radiation in Super-Rosenbluth experiments made in Ref.~\cite{Qattan:2004ht}. First, we repeat the approach used in Ref.~\cite{Qattan:2004ht} based on a Monte Carlo simulation of radiative effects in the peaking approximation, and we confirm that the radiative corrections are in fact smaller and less $\varepsilon$-dependent than a corresponding electron-detection experiment. Then, we examine a more sophisticated model of radiative effects that go beyond the peaking approximation. We find that the more sophisticated model has enormous variance in event weights due to dramatic rise in the cross section for hard radiation emitted from the proton leg in the direction of scattered electron, which is not captured in the peaking approximation. We conclude that the peaking approximation may be misleadingly inaccurate for Super Rosenbluth experiments and see a need for numerically-stable non-peaking approaches for future experiments.

\section{Methodology}
\label{sec:methods}

We have evaluated the radiative correction prescribed by different models of radiative effects by performing Monte Carlo pseudo-experiments. Traditionally, radiative corrections have been formulated in terms of a correction, $\delta$, such that:
\begin{equation}
    d\sigma_\text{rad.} = (1+\delta) d \sigma_\text{Born},
    \label{eq:nonexp}
\end{equation}
where $d\sigma_\text{rad.}$ is the measured cross section which includes radiative effects and $d \sigma_\text{Born}$ is the corresponding cross section under the Born approximation, i.e., one-photon exchange only. Various prescriptions for $\delta$ have been suggested (see, e.g., \cite{Tsai:1961zz,Meister:1963zz,Mo:1968cg,Maximon:2000hm}), which account for effects at the next-to-leading-order level, including vertex corrections, two-photon exchange, vacuum polarization, and the emission of a soft bremsstrahlung photon, which is experimentally indistinguishable from elastic scattering. It was pointed out, first in Ref.~\cite{Yennie:1961ad}, that accounting for soft-bremsstrahlung to all orders can be 
approximated by exponentiating the correction, i.e., 
\begin{equation}
    d\sigma_\text{rad.} = e^\delta d \sigma_\text{Born},
    \label{eq:exp}
\end{equation}
As we consider both exponentiated and non-exponentiated models in this paper, we choose to define the radiative correction factor $f_\text{RC}$ as the ratio of the cross section including radiative effects to the equivalent cross section under the Born approximation:
\begin{equation}
    f_\text{RC} \equiv \frac{d\sigma_\text{rad.}}{d\sigma_\text{Born}}.
\end{equation}

We consider only the elastic electron-proton scattering reaction in the fixed-target frame, in which the final state has a scattered electron, 
with scattering angle $\theta_e$ and outgoing momentum $p_e$, a recoiling proton emerging at an angle $\theta_p$ and momentum $p_p$, and possibly 
some energy and momentum radiated in the form of photons. In this reaction, $Q^2$ can be determined from the outgoing lepton kinematics, i.e.,
\begin{equation}
    Q^2 = 2 E_\text{beam} E_e - 2 p_\text{beam} p_e \cos\theta_e - 2m_e^2,
\end{equation}
    with $E_\text{beam}$ and $p_\text{beam}$ being the incoming beam energy and momentum, respectively, $E_e$ being the energy of the outgoing electron, and $m_e$ being the lepton mass. Alternatively, $Q^2$ can be defined from the outgoing proton kinematics, i.e.,
 \begin{equation}
    Q^2 = 2E_p m_p - 2m_p^2
\end{equation}   
    where $E_p$ is the outgoing proton energy and $m_p$ is the proton mass. In the absence of any energy radiated by photons, these definitions coincide.
Another important variable considered in Rosenbluth separations is $\varepsilon$, the virtual photon polarization parameter, which, in terms of electron kinematics, is defined as:
\begin{equation}
    \varepsilon = \left[ 1+ 2(1+\tau)\tan^2\frac{\theta_e}{2}\right]^{-1}
\end{equation}
    where $\tau \equiv Q^2/4m_p^2$. 

In order to calculate $f_\text{RC}$, we ran Monte Carlo simulations using different event generators with different models for incorporating radiative effects. Simulations were conducted in the form of pseudo-experiments; after generating samples of Monte Carlo events, events were accepted if they fell within an acceptance window around a central kinematic point. We used a polar angle ($\theta$) acceptance of $\pm 1$~mrad, small enough to heavily suppress effects due to the change in cross section across the acceptance window. 
This is substantially smaller than the acceptance of most high-resolution spectrometers. For example, a Hall A High Resolution Spectrometer (HRS), used to make the measurement in Ref.~\cite{Qattan:2004ht}, has a horizontal angle acceptance of approximately $\pm 25$ mrad, depending on the desired momentum acceptance~\cite{Alcorn:2004sb}. Our choice was motivated by the goal of evaluating the radiative correction at specific kinematic points, rather than realistically simulating any specific experiment. Since the unpolarized cross section has no azimuthal dependence, i.e., is azimuthally symmetric, the size of the azimuthal acceptance window has no impact on our results. We therefore assumed a full $2\pi$ azimuthal acceptance. 

\begin{table}[]
    \centering
    \caption{Kinematic points studied}
    \begin{tabular}{c c c c c c c}
    \hline\hline
         $Q^2$ & $\varepsilon$ & $E_\text{beam}$ [GeV] & $\theta_e$ [$^\circ$] & $p_e$ [GeV$/c$] & $\theta_p$ [$^\circ$] & $p_p$ [GeV$/c$] \\
         \hline
     & 0.1 & 1.891 & 116.11 & 0.485 & 11.68 & 2.149 \\
     & 0.3 & 2.168 & 78.47 & 0.761 & 20.30 & 2.149 \\
2.64 & 0.5 & 2.565 & 56.26 & 1.158 & 26.61 & 2.149 \\
     & 0.7 & 3.262 & 38.58 & 1.855 & 32.55 & 2.149 \\
     & 0.9 & 5.388 & 20.21 & 3.981 & 39.77 & 2.149 \\
     \hline
    & 0.1 & 2.219 & 113.85 & 0.514 & 10.95 & 2.471 \\
    & 0.3 & 2.537 & 76.04 & 0.831 & 19.05 & 2.471 \\
3.2 & 0.5 & 2.993 & 54.21 & 1.288 & 25.00 & 2.471 \\
    & 0.7 & 3.794 & 37.05 & 2.089 & 30.61 & 2.471 \\
    & 0.9 & 6.239 & 19.36 & 4.534 & 37.46 & 2.471 \\
    \hline
    & 0.1 & 2.739 & 110.52 & 0.554 & 10.03 & 2.979 \\
    & 0.3 & 3.122 & 72.57  & 0.937 & 17.47 & 2.979 \\
4.1 & 0.5 & 3.672 & 51.34  & 1.487 & 22.95 & 2.979 \\
    & 0.7 & 4.638 & 34.93  & 2.453 & 28.14 & 2.979 \\
    & 0.9 & 7.585 & 18.20  & 5.400 & 34.49 & 2.979 \\
    \hline
    \end{tabular}
    \label{tab:kin}
\end{table}

For every kinematic point we studied, we conducted two pseudo-experiments: one in which the spectrometer was positioned to detect the electron, and another in which the spectrometer was positioned to detect the proton, i.e., a Super-Rosenbluth configuration. For each kinematic point we studied, we performed separate simulation runs. We studied 15 kinematic points in total. We considered the three $Q^2$ values of the data in Ref.~\cite{Qattan:2004ht}: 2.64, 3.2, and 4.1 GeV$^2/c^2$, and at each we tested five values of $\varepsilon$ from 0.1 to 0.9, opting for a wider $\varepsilon$-coverage, rather than exactly matching the kinematic points reported in Ref.~\cite{Qattan:2004ht}. The elastic reaction kinematics for these points are given in Table \ref{tab:kin}.

To avoid any needless complexity, for all Monte Carlo codes, the proton is assumed to have dipole form factors, i.e.,
\begin{align}
    G_E =& \left[1 + \frac{Q^2}{\Lambda^2}\right]^{-2},\\
    G_M =& \mu_p G_E,
\end{align}
    with $\Lambda = 0.71$~GeV$^2/c^2$ and $\mu_p=2.79$. The dipole model has well known deficiencies (see, for example, \cite{A1:2013fsc}), but we want to avoid the appearance that some specific form factor model is germane or even responsible for our findings. 

\section{Peaking Approximation Results}
\label{sec:results}

The first radiative correction model we considered was the pure-peaking approximation described by Ent et al., developed for the NE-18 Experiment~\cite{Ent:2001hm}. This approach is widely used and incorporated in many standard simulation codes, including \texttt{simc}~\cite{SIMC}. 
This approach was used to determine the radiative corrections in the experiment of Ref.~\cite{Qattan:2004ht}. In this approach, radiation
is assumed to be emitted independently from the incoming electron, the outgoing electron, and the outgoing proton, purely in the direction of
the particle momentum vectors, i.e., the incoming electron radiates photons in the direction of the beam, the outgoing electron radiates in the direction $\theta_e$, and the outgoing proton radiates in the direction $\theta_p$. Each radiating particle can emit multiple photons; the only relevant quantity is the total energy radiated from the particle. The cross section for the scattering reaction is evaluated with an updated beam energy, i.e., the electron energy remaining after emitting initial state radiation. The effect of loop diagrams (i.e. vacuum polarization, vertex corrections), which are non-exponentiated, are also evaluated in the updated kinematics. The radiative cross section is given in Eq.~66 of Ref.~\cite{Ent:2001hm}. 

The peaking approximation is often employed because the angular distribution of bremsstrahlung radiation is sharply peaked in the directions of the momentum vectors of the incoming and outgoing particles, especially those of the electron, due to its low mass. However, because radiation from each particle leg is treated independently, this approach is inaccurate in its treatment of interference terms, e.g. the interference of bremsstrahlung emitted from the incoming electron with bremsstrahlung emitted from the outgoing electron.

\begin{figure*}[h]%
\centering
\includegraphics[width=0.46\textwidth]{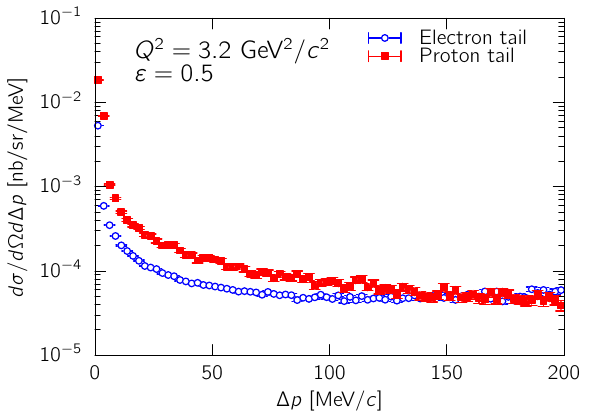}\hspace{0.05\textwidth}
\includegraphics[width=0.46\textwidth]{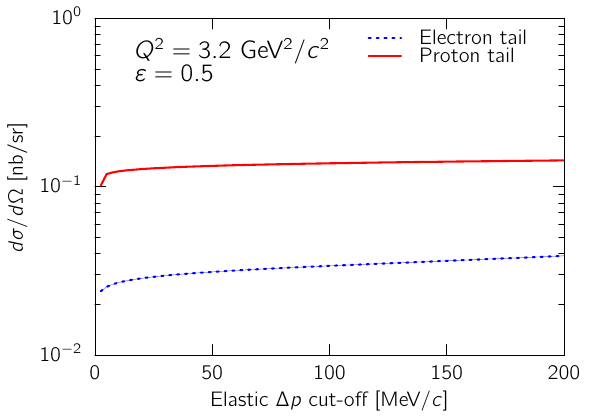}
\caption{Left: an example radiative tail from the pure-peaking (NE-18 prescription~\cite{Ent:2001hm}) Monte Carlo generator for electron detection (blue) vs.\ proton detection (red). 
Right: the resulting integral of the radiative tail over the elastic region, i.e., from 0 up to some cut-off in $\Delta p$, as a function of that cut-off. The large difference in normalization between the two tails comes primarily from the Jacobian between the electron solid angle and proton solid angle.  
\label{fig:tailshape}}
\end{figure*}

Example radiative tails are shown in Fig.~\ref{fig:tailshape} for a pseudo experiment conducted at $Q^2 = 3.2$~GeV$^2/c^2$ and $\varepsilon=0.5$, which corresponds to the elastic kinematics of $E_\text{beam}=2.993$~GeV, $\theta_e = 54.21^\circ$, $p_e=1.288$~GeV$/c$, $\theta_p=25.00^\circ$, and $p_p=2.471$~GeV$/c$. In both panels, the data in blue correspond to an electron-detection experiment, while the data in red correspond to an experiment in which only the proton is detected. The left panel shows the momentum spectrum that would be measured by a spectrometer in terms of $\Delta p$, the momentum loss due to radiation. We define this momentum loss as the magnitude of the difference between its momentum and the momentum a particle would have at that angle absent any radiation:
\begin{equation}
\Delta p \equiv
    \begin{cases}
        \frac{E_\text{beam} m_p}{m_p + E_\text{beam} (1-\cos\theta_e)} - p_e & \text{Electron detection}\\
        &\\
        \frac{2m_pE_\text{beam}(E_\text{beam} +m_p)\cos\theta_p}{E_\text{beam}^2\sin^2\theta_p + 2E_\text{beam}m_p + m_p^2} - p_p
        & \text{Proton detection}
    \end{cases}
    \label{eq:DeltaP}
\end{equation}
(neglecting the electron mass). The minimum $\Delta p$ is zero, and increasing $\Delta p$ means detecting a particle with less and less momentum. One should note that the momentum loss need not come exclusively from final state radiation from the detected particle. There can also be initial state radiation as well, reducing the effective beam energy, leading to a lower momentum scattered particle. The right panel of Fig.~\ref{fig:tailshape} shows the integrated elastic cross section as a function of the cut-off point one uses to define which events are sufficiently elastic. At that kinematic point, the differential cross section $d\sigma/d\Omega$ in the proton detection experiment is approximately four times larger than that in the electron detection experiment due to the Jacobian between $\Omega_e$ and $\Omega_p$. The tails at other kinematic points were qualitatively similar to those shown in Fig.~\ref{fig:tailshape}.

\begin{figure*}[h]%
\centering
\includegraphics[width=0.46\textwidth]{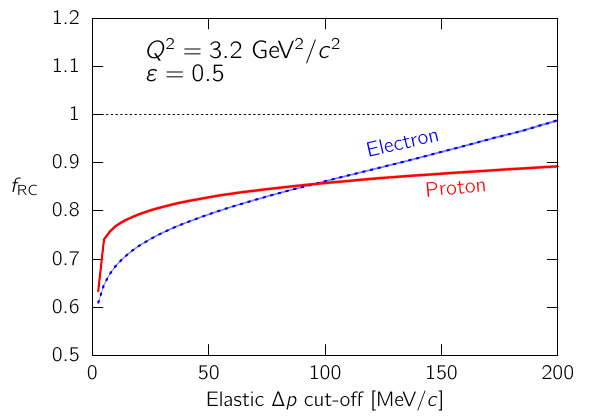}\hspace{0.05\textwidth}
\includegraphics[width=0.46\textwidth]{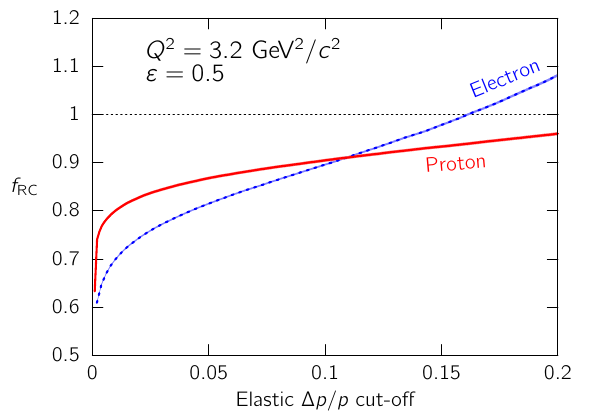}
\caption{The radiative correction corresponding to the pure-peaking tails in Fig.~\ref{fig:tailshape} as a function of the $\Delta p$ momentum cut-off for electron detection (blue) and proton detection (red). The left plot shows the correction for an absolute $\Delta p$, while the right plot gives the correction in terms of the relative cut-off, $\Delta p/p$, which more closely corresponds to the acceptance of a magnetic spectrometer. }
\label{fig:frc}
\end{figure*}

By comparing the cross sections in the right panel of Fig.~\ref{fig:tailshape} to the Born cross section, we can determine the radiative correction factor $f_\text{RC}$, shown in Fig.~\ref{fig:frc}, for the same kinematics as Fig.~\ref{fig:tailshape}. The left panel shows $f_\text{RC}$ as a function of the $\Delta p$ cut-off. The right panel shows the same correction factor plotted against a cut off point in $\Delta p/p$, i.e, the relative change in momentum. This is arguably more representative of experiments using dipole magnetic spectrometers, whose momentum acceptances typically cover a fraction of the central momentum. It can be seen in both panels that in the proton detection experiment, $f_\text{RC}$ has a smaller slope with respect to the cut-off point. The behavior of the radiative corrections are qualitatively similar at the other kinematic points we studied.

\begin{figure*}[h]%
\centering
\includegraphics[width=0.46\textwidth]{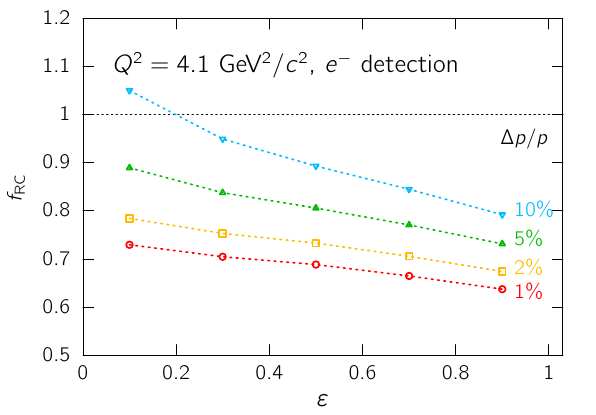}\hspace{0.05\textwidth}
\includegraphics[width=0.46\textwidth]{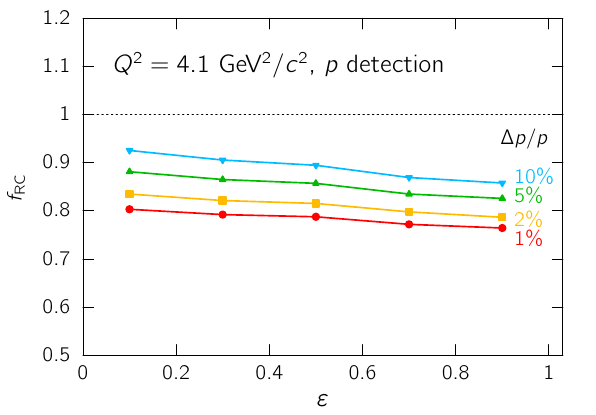}
\caption{The radiative correction factors predicted by the NE18 prescription~\cite{Ent:2001hm} of the peaking approximation at $Q^2=4.1$~GeV$^2/c^2$
for a traditional electron detection experiment (left panel) versus a proton detection experiment (right panel), as a function of several different
elastic cut-off momenta. In the proton-detection experiment, the radiative correction factors have less $\varepsilon$-dependence.}
\label{fig:eps_dep}
\end{figure*}

Fig.~\ref{fig:eps_dep} shows the radiative correction factor evaluated for a range of $\varepsilon$ and relative momentum cut-off values for electron detection (left) and proton detection (right) experiments. The figure shows the results for the $Q^2=4.1$~GeV$^2/c^2$ kinematics, but the results are qualitatively similar for all three $Q^2$ values. The radiative correction factors in the proton detection experiment have reduced dependence on $\varepsilon$ compared to those in an equivalent electron detection experiment. Furthermore, at higher values of $\varepsilon$, the proton-detection $f_\text{RC}$ is closer to unity. This result validates the claims of Ref.~\cite{Qattan:2004ht}. Within the peaking approximation, radiative corrections in proton detection appear to be less likely to pose a risk of bias on a Rosenbluth separation. 

\section{Non-Peaking Results}

The suitability of the peaking approximation for Super-Rosenbluth experiments can be validated by calculations using radiative models that do not make the peaking approximation. We attempted calculations with two such models. The first, ESEPP~\cite{Gramolin:2014pva}, was developed with particular attention to exactly calculating the bremsstrahlung process at lowest order, even preserving terms with the electron mass. This is critical for high-precision experiments at lower energies or lower values of $Q^2$, such as those attempting to determine the proton radius. ESEPP does not treat multi-photon bremsstrahlung. However, its treatment of single-photon bremsstrahlung is exact, up to knowledge of proton structure, including all interference terms. Unfortunately, ESEPP, in its current form, is unable to perform a calculation for a Super Rosenbluth experiment. Our attempts to run the generator over the full electron phase space caused the program to crash, a problem we were unable to solve.

The second model we considered was the radiative event generator developed for the OLYMPUS Experiment~\cite{OLYMPUS:2013lem}. The OLYMPUS Generator (described so far in Refs.~\cite{Schmidt:2016fsl,Russell:2016syx}, with an article in preparation), treats multi-photon bremsstrahlung with an exponentiation approach and avoids the peaking approximation, instead evaluating the bremsstrahlung matrix element including all interference terms. This approach was important for the primary goal of the OLYMPUS Experiment: measuring any deviations from unity in the ratio of positron-proton to electron-proton elastic scattering cross sections~\cite{OLYMPUS:2016gso}. The interference of lepton- and proton-bremsshtralung produces an asymmetry between electron and positron cross sections and was thus a background for the measurement~\cite{OLYMPUS:2016gso}. 

The OLYMPUS Generator produces events with the kinematics of one-photon emission, i.e., $ep\rightarrow ep\gamma$, by randomly sampling the electron angles 
$\theta_e$ and $\phi_e$, the momentum loss of the electron, $\Delta p_e$ (as in Eq.~\ref{eq:DeltaP}), and angles of the radiated photon. The proton momentum vector is determined from energy and momentum conservation. The events are weighted in order to achieve a model cross section of the form:
\begin{equation}
    \frac{d^5\sigma}{d\Omega_e d\Delta p_e d\Omega_\gamma } = \frac{d^5\sigma}{d\Omega_e d\Delta p_e d\Omega_\gamma }_\text{brems.} \times e^{\delta(\theta_e,\phi_e,\Delta p_e)},
\end{equation}
where the $d^5\sigma_\text{brems.}$ term is the tree-level bremsstrahlung cross section, and $\delta(\theta_e,\phi_e,\Delta p_e)$ is a standard radiative correction such as appears in Eqs.~\ref{eq:nonexp} and \ref{eq:exp}. The sampling distributions for the kinematic variables were designed to approximate the bremsstrahlung cross section in order to reduce the variance in event weights. See Ref.~\cite{Schmidt:2016fsl} for further details. In this approach, the radiated photon is not associated with any of the charged particles in the reaction. When calculating, $d^5\sigma_\text{brems.}$ for an event, four diagrams---photon emission from the initial state electron, final state electron, initial state proton, and final state proton---all contribute coherently. The diagrams with emission from the electron legs are also called the Bethe-Heitler process~\cite{Bethe:1934za}. The diagrams with emission from the proton legs can be referred to as virtual Compton scattering (VCS). However, both processes contribute coherently in each event. 
It should be noted that in the evaluation of the diagrams, the proton vertices are evaluated with on-shell form factors. It should also be noted that since the OLYMPUS Experiment used a thin gas target that produced negligible external radiation, the OLYMPUS generator has no treatment of external bremsstrahlung. 

\begin{figure*}[h]%
\centering
\includegraphics[width=0.66\textwidth]{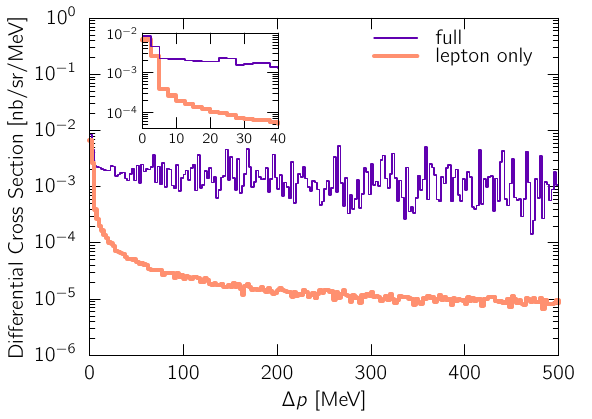}
\caption{Radiative tail prediction for a proton-detection experiment using the OLYMPUS radiative event generator, for the $\varepsilon=0.5$, $Q^2=3.2$~GeV$^2/c^2$ kinematic point. 
The full calculation includes contributions from all bremsstrahlung diagrams, while the lepton-only calculation used a modified version of the OLYMPUS generator that neglects diagrams in which bremsstrahlung photons are radiated from the proton. Radiation from the proton causes the OLYMPUS generator
to have large fluctuations in event weights when integrating over the full electron phase space.}
\label{fig:olympus}
\end{figure*}

The OLYMPUS Generator was able to run when full electron phase space was requested.
The radiative tail produced by running the OLYMPUS generator for a proton detection experiment at $Q^2=3.2$~GeV$^2/c^2$ and $\varepsilon=0.5$ is shown in Fig.~\ref{fig:olympus}. The results, when running the full bremsstrahlung calculation (shown by the thin purple histogram), have large fluctuations in event weights. This indicates a possible divergence or other unregulated numerical problem when trying to cover the full electron phase space. By contrast, when running a modified version of the generator in which the bremsstrahlung calculation was limited to only the two diagrams in which the electron radiates (shown by the thicker orange histogram, labelled ``lepton only''), the variance is eliminated. Even though the full electron phase space is being integrated over, it is radiation from the proton that is causing the variance problem within the generator. 

\begin{figure*}[h]%
\centering
\includegraphics[width=0.48\textwidth]{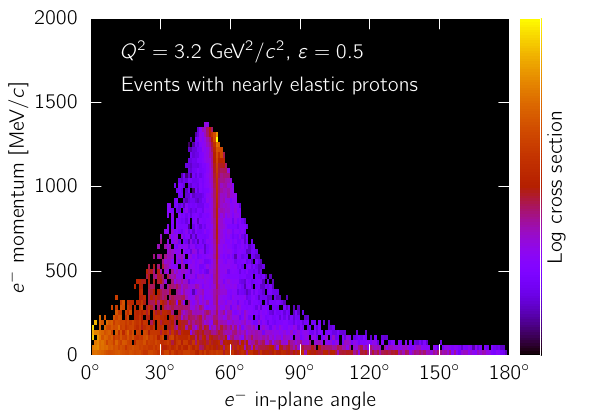} 
\includegraphics[width=0.48\textwidth]{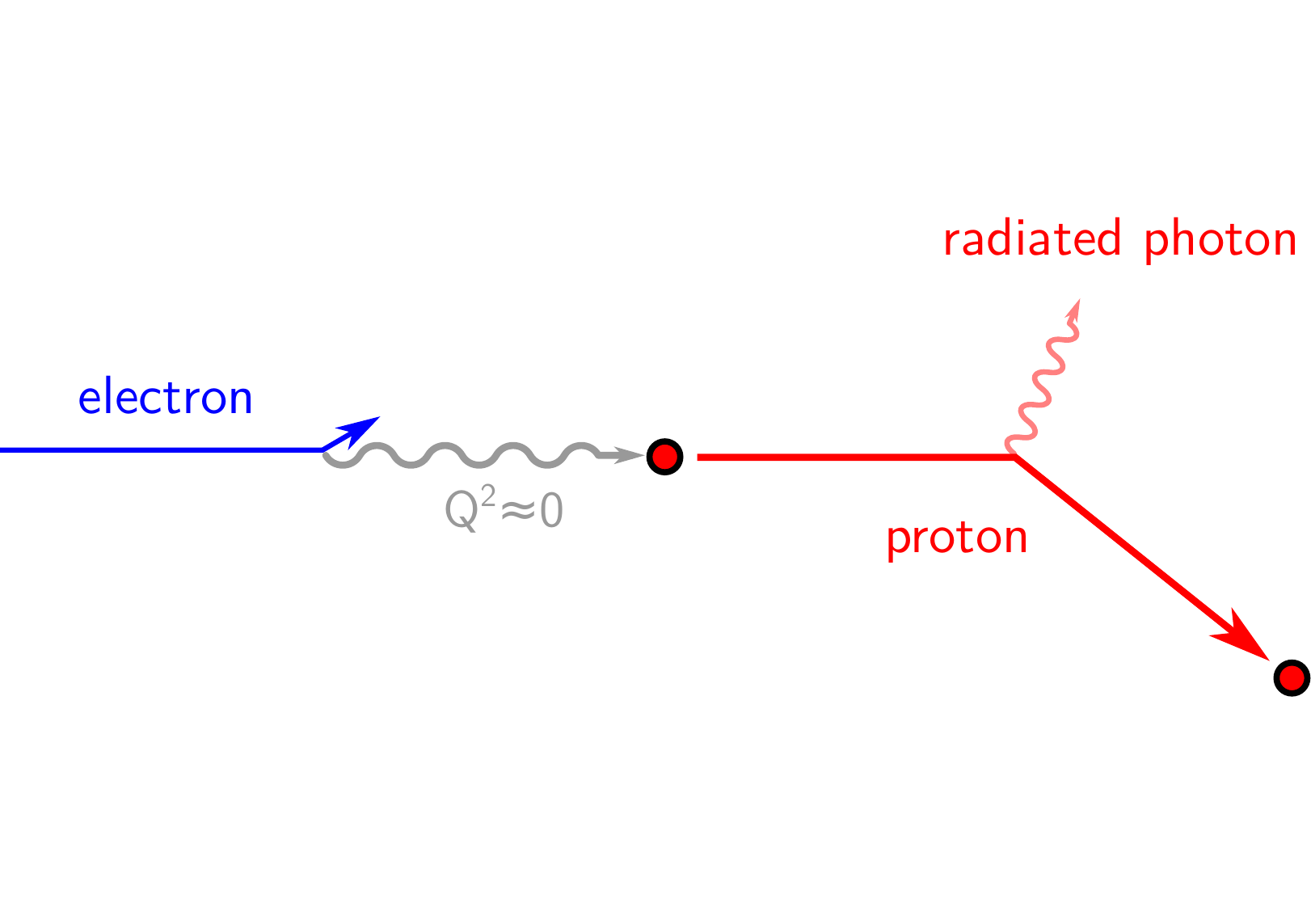}
\caption{Left: the electron energy and scattering angle for events in our proton detection pseudo experiment (using the OLYMPUS radiative generator) in which the proton is detected as nearly elastic, i.e., within 5\% of the elastic momentum given its angle. There is a large enhancement in cross section for low-energy electrons. Right: a schematic of how this enhancement occurs. The electron can produce a virtual photon with $Q^2=0$, but with nearly the full beam energy. This virtual photon can Compton scatter from the target proton, giving it an angle and momentum similar to those of elastic scattering.} 
\label{fig:trouble}
\end{figure*}

To examine this further, we considered events in which the proton would have been measured in the spectrometer as being ``nearly elastic,'' i.e., within 5\% of the elastic momentum given the proton's angle, and looked where the electrons were going. The results of this study (conducted at $Q^2=3.2$~GeV$^2/c^2$ and $\varepsilon=0.5$) are shown in the left panel of Fig.~\ref{fig:trouble}. The expected elastic peak can be found at approximately 54$^\circ$ and 1300~MeV$/c$. However, there is also a large increase in cross section for events in which the electron has lost nearly all of its momentum. The radiated bremsstrahlung photon in such events is primarily emitted from the proton but in the direction of otherwise elastic electrons, i.e., 54$^\circ$. A cartoon of the process that leads to this enhancement is shown on the right side of Fig.~\ref{fig:trouble}. In losing all of its energy, the incoming electron produces a virtual photon with nearly the same energy and direction as the beam. This virtual photon can Compton scatter from the proton, the kinematics of which are nearly identical to elastic scattering. The enhancement in cross section comes from the virtual photon having $Q^2 \approx 0$. The virtual photon propagator carries a factor of $1/Q^2$. 

It appears that this process does not lead to any divergences ($Q^2$ is never exactly zero), but the enhancement in the cross section is not well matched by the sampling distribution, leading to large variance in event weights. Large weight variance by itself is not necessarily cause for alarm; it merely slows the convergence of any Monte Carlo calculation. However, large weight variance at a kinematic end point suggests that the generator might be vulnerable to numerical problems caused by rounding at this end point. 
Given how ESEPP evaluates the bremsstrahlung matrix elements, it is likely to suffer from similar problems. 
The numerical problems in the OLYMPUS generator could be repaired by modifying the sampling distribution from which kinematic variables are randomly chosen, so that the rapid changes in cross section are matched by similar changes in the sampling distribution. We think this approach is worth further investigation.

Alarmingly, this process is completely neglected in the peaking approximation. The peaking approximation allows radiation to be emitted from the outgoing proton; however, this radiation is restricted to being purely in the direction of the outgoing proton. There is no mechanism within this approximation to handle proton radiation in the elastic lepton direction. 
Radiation from the proton is also neglected in another previous non-peaking calculation of radiative corrections in proton detection experiments~\cite{Afanasev:2001nn}. 

\begin{figure}[htpb]
    \centering
    \includegraphics[width=10cm]{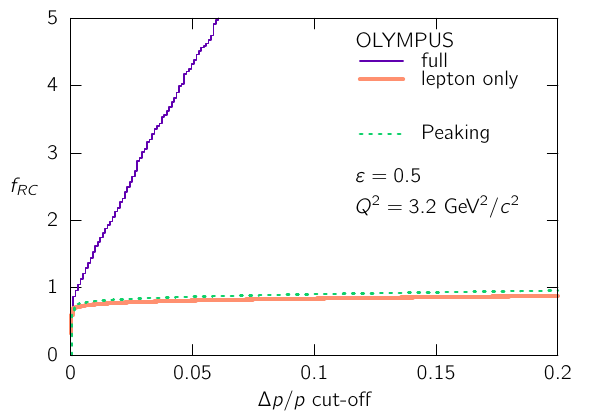}
    \caption{
        \label{fig:olympus_frc}
    A comparison of the radiative correction factor, $f_\text{RC}$ for a proton-detection experiment as predicted by the full OLYMPUS generator, the modified OLYMPUS generator that only considers radiation from lepton legs, and the generator using the NE18 peaking approximation~\cite{Ent:2001hm}. The kinematics are for $Q^2=3.2$~GeV$^2/c^2$ and $\varepsilon=0.5$, in which the proton has an elastic momentum of 2.471~GeV$/c$. Note that the peaking approximation results closely match those of the OLYMPUS lepton-only calculation. }
\end{figure}

The impact of proton radiation can be seen by integrating the tails in Fig.~\ref{fig:olympus} up to some $\Delta p$ cut-off to produce the radiative correction factor $f_\text{RC}$, which is shown in Fig.~\ref{fig:olympus_frc}. For comparison, we also include the $f_\text{RC}$ predicted using the peaking approximation (shown earlier in Fig.~\ref{fig:frc}). The peaking result is very close to that of the modified OLYMPUS generator that only considers radiation from the electron legs (lepton only). The result of the full OLYMPUS generator very quickly deviates and predicts an enormous enhancement to the cross section. 

There are several reasons to treat the full OLYMPUS generator calculation with skepticism. As mentioned above, the large variance in event weights calls into question how well the OLYMPUS generator handles numerics at the kinematic end point where the out-going electron has zero energy. Next, the OLYMPUS generator evaluates the proton vertices in the bremsstrahlung diagrams using on-shell proton form factors. A more detailed description using Compton Form Factors (CFFs) would be more accurate. Finally, the results of the Super Rosenbluth experiment presented in Ref.~\cite{Qattan:2004ht} are generally in line with prior electron-proton scattering measurements, i.e, no colossal enhancement in the cross section was observed. The $\Delta p$ spectra shown in Ref.~\cite{Qattan:2004ht} extend to approximately $\pm 1\%$, i.e., smaller than the range shown in Fig.~\ref{fig:olympus_frc}, so it is not clear what happens to the radiative tail as $\Delta p$ extends out to 10\%--20\%. It should also be noted that in Ref.~\cite{Qattan:2004ht}, there is a subtraction of background coming from Compton scattering, presumed to be from photons produced by the beam hitting upstream material, such as target windows. This subtracted background is not independently normalized; instead, the normalization is estimated from the measured spectrum. This means that any radiative tail contributions from proton radiation (i.e., internal virtual Compton scattering) may be partially (or even totally) accounted by the procedure to subtract external, real Compton scattering. 

\section{Conclusions}
\label{sec:conclusions}

We have investigated radiative corrections to elastic electron-proton scattering for Super Rosenbluth experiments, in which the recoiling proton is detected instead of the scattered electron.
The Super Rosenbluth technique has several experimental advantages, and more favorable radiative corrections has been put forward as one of them~\cite{Qattan:2004ht}. 
We performed Monte Carlo pseudo-experiments using two different radiative models to evaluate that claim. 
When using the peaking approximation, we find that the radiative corrections are indeed less $\varepsilon$-dependent than in equivalent electron-detection experiments, confirming the claims of Ref.~\cite{Qattan:2004ht}. However, in attempting to use a non-peaking model, we encountered numerical instabilities, which we attribute to a large enhancement in the cross section near $Q^2 \approx 0$ when the proton radiates a bremsstrahlung photon in the elastic electron direction. 
We see no reason why this process would not contribute to an actual proton-detection experiment. 
Nevertheless, the process is neglected in the peaking approximation, which raises questions about the suitability of the peaking approximation for Super Rosenbluth experiments. 

One argument given for the favorability of radiative corrections in the Super Rosenbluth technique is that one detects the proton, and the proton radiates to a much less degree than the lighter electron. This is true. However, the evaluation of radiative corrections depends as much on the particles being detected as the particles that are undetected. By not detecting the electron, the evaluation of the radiative correction requires integrating over all of the possible phase space where the electron can end up. In that respect, evaluating the radiative correction in electron detection experiments may be an easier problem. 

Given the other experimental benefits of the Super Rosenbluth technique, developing a robust radiative corrections procedure that avoids reliance on the peaking approximation should be pursued. None of the current Monte Carlo event generators we investigated are up to the task. 

\bmhead{Acknowledgments}

This work was supported by the US Department of Energy Office of Science, Office of Nuclear Physics, under contract no. DE-SC0016583.
Q.~S.\ acknowledges support from a Luther Rice Fellowship from the George Washington University's Columbian College of Arts and Sciences
and from a Walker Fellowship from the Department of Physics. 

\bibliography{sn-bibliography}

\end{document}